\documentclass[aps,pre,twocolumn,superscriptaddress,showpacs,preprintnumbers]{revtex4-1}
\usepackage[colorlinks=true, pdfstartview=FitV, linkcolor=red, citecolor=blue, urlcolor=black, pdftitle={},pdfauthor={Tomoya Hayata},pdfsubject={}, pdfkeywords={}]{hyperref}
\usepackage{graphicx}
\usepackage{setspace,bm}
\usepackage{latexsym,amssymb,amsmath,mathrsfs}
\usepackage{color}
\graphicspath{{./figure/}}

\makeatletter
\providecommand \@ifxundefined [1]{%
 \@ifx{#1\undefined}
}%
\providecommand \@ifnum [1]{%
 \ifnum #1\expandafter \@firstoftwo
 \else \expandafter \@secondoftwo
 \fi
}%
\providecommand \@ifx [1]{%
 \ifx #1\expandafter \@firstoftwo
 \else \expandafter \@secondoftwo
 \fi
}%
\providecommand \bibnamefont  [1]{#1}%
\providecommand \bibfnamefont [1]{#1}%
\providecommand \citenamefont [1]{#1}%
\providecommand \href@noop [0]{\@secondoftwo}%
\providecommand \href [0]{\begingroup \@sanitize@url \@href}%
\providecommand \@href[1]{\@@startlink{#1}\@@href}%
\providecommand \@@href[1]{\endgroup#1\@@endlink}%
\providecommand \@sanitize@url [0]{\catcode `\\12\catcode `\$12\catcode
  `\&12\catcode `\#12\catcode `\^12\catcode `\_12\catcode `\%12\relax}%
\providecommand \@@startlink[1]{}%
\providecommand \@@endlink[0]{}%
\providecommand \url  [0]{\begingroup\@sanitize@url \@url }%
\providecommand \@url [1]{\endgroup\@href {#1}{\urlprefix }}%
\providecommand \urlprefix  [0]{URL }%
\providecommand \Eprint [0]{\href }%
\providecommand \doibase [0]{http://dx.doi.org/}%
\providecommand \selectlanguage [0]{\@gobble}%
\providecommand \bibinfo  [0]{\@secondoftwo}%
\providecommand \bibfield  [0]{\@secondoftwo}%
\providecommand \BibitemOpen [0]{}%
\providecommand \EOS [0]{\spacefactor3000\relax}%
\providecommand \BibitemShut  [1]{\csname bibitem#1\endcsname}%
\let\auto@bib@innerbib\@empty

\newcommand{\vphi}{\varphi}
\newcommand{\be}{\begin{equation}}      
\newcommand{\ee}{\end{equation}}      
\newcommand{\bea}{\begin{eqnarray}}      
\newcommand{\eea}{\end{eqnarray}}

\begin{document}
\title{Chiral magnetic effect of light}

\author{Tomoya Hayata}
\affiliation{
Department of Physics, Chuo University, 1-13-27 Kasuga, Bunkyo, Tokyo, 112-8551, Japan 
}

\date{\today}

\begin{abstract}

We study a photonic analog of the chiral magnetic (vortical) effect. 
We discuss that the vector component of magnetoelectric tensors plays a role of ``vector potential," 
and its rotation is understood as ``magnetic field" of a light. 
Using the geometrical optics approximation, 
we show that ``magnetic fields" cause an anomalous shift of a wave packet of a light through an interplay with the Berry curvature of photons. 
The mechanism is the same as that of the chiral magnetic (vortical) effect of a chiral fermion, 
so that we term the anomalous shift ``chiral magnetic effect of a light."
We further study the chiral magnetic effect of a light beyond geometric optics by directly solving the transmission problem of a wave packet at a surface of a magnetoelectric material.
We show that the experimental signal of the chiral magnetic effect of a light is the nonvanishing of transverse displacements for the beam normally incident to a magnetoelectric material.

\end{abstract}

\maketitle

\section{Introduction} 
The Berry phase~\cite{Berry45} and Berry curvature have attracted a lot of interests in subdisciplines of physics such as condensed matter physics, nuclear physics, and particle physics.
They characterize topology of wave functions in momentum space and explain many properties of topological materials such as topological insulators~\cite{PhysRevB.78.195424,RevModPhys.82.3045}, topological superconductors~\cite{RevModPhys.83.1057}, and Weyl/Dirac semimetals~\cite{2017arXiv170207310Z,2017arXiv170501111A} as well as those of relativistic Weyl/Dirac fermions such as quarks and neutrinos. 

It has been known that photons have a nonzero Berry curvature, which originates from the massless and helical nature of them~\cite{PhysRevLett.57.933,PhysRevLett.57.937,Berry87}.
The Berry curvature of photons leads to novel effects, which cannot be explained by the standard geometric optics according to Fermat's principle.
The famous example is the Hall effect of light~\cite{PhysRevLett.93.083901,PhysRevE.74.066610}, which is also known as the optical Magnus effect~\cite{PhysRevA.45.8204,PhysRevA.46.5199,PhysRevD.74.021701,DUVAL2007925}.
A spatially varying refractive index can be understood as an ``electric field" of a light, and causes transverse shifts of a wave packet of a light. 
This deviation from Snell's law in a finite beam has been confirmed by experiments~\cite{PhysRevLett.96.073903,Hosten787}. 
The phenomena can be explained by considering the anomalous group velocity of a light due to the interplay between ``electric fields" and Berry curvature, 
and the mechanism is the same as that of the anomalous Hall effect in electron systems~\cite{RevModPhys.82.1539,RevModPhys.82.1959}. 

Such an analogy between geometric optics and semiclassical dynamics of electrons is not complete yet. 
This is because photons do not couple with magnetic fields, and no Lorentz force appears in the standard geometric optics. 
So far, it has been indicated to use inhomogeneous optical magnetoelectric effect as ``magnetic fields" of a light in Ref.~\cite{PhysRevLett.95.237402}.
There have also been works to study the physics of photons under ``magnetic fields" in the context of photonic crystals~\cite{PhysRevLett.100.013904,PhysRevA.78.033834,PhysRevLett.100.013905,Hafezi2011,Fang2012,Hafezi2013,Lu2014}.
However a magnetic analog of anomalous Hall effect, that is, the exotic phenomenon originated from the interplay between ``magnetic fields" and Berry curvature~\cite{PhysRevLett.109.162001,PhysRevLett.109.181602,PhysRevLett.110.262301,PhysRevB.88.104412},
which is referred to as the chiral magnetic (vortical) effect in the physics of chiral fermions~\cite{PhysRevD.78.074033,PhysRevD.20.1807,PhysRevLett.103.191601}, 
has not been discussed.

In this paper, we study a photonic analog of the chiral magnetic (vortical) effect~\cite{PhysRevD.78.074033,PhysRevD.20.1807,PhysRevLett.103.191601}.
We discuss that rotation of the vector component of magnetoelectric tensors behaves as ``magnetic fields" of a light~\cite{PhysRevLett.95.237402},  
and causes anomalous shifts of a wave packet of a light along the direction parallel to it. 
In terms of geometrical optics,  the mechanism is the same as that of the chiral magnetic (vortical) effect of chiral fermions~\cite{PhysRevLett.109.162001,PhysRevLett.109.181602,PhysRevLett.110.262301,PhysRevB.88.104412}, and thus we term the helicity-dependent shifts ``chiral magnetic effect of a light."
We show that  the helical shifts due to the chiral magnetic effect of a light arise beyond the geometrical optics approximation.
We also discuss an analog of the spectral flow~\cite{PhysRevB.88.104412,NIELSEN1983389} in geometric optics.

The same effect arising in rotating frame was discussed in Refs.~\cite{2017PhRvA..96d3830Z,Avkhadiev:2017fxj,Yamamoto:2017uul}. 
References~\cite{2017PhRvA..96d3830Z,Avkhadiev:2017fxj} calculated helicity current of photons only in equilibrium states at finite temperature.
Reference~\cite{Yamamoto:2017uul} calculated helicity current using the kinetic theory with the Berry curvature correction,
which is, in principle, applicable to nonequilibrium states.
However the wave packet dynamics was not studied in Ref.~\cite{Yamamoto:2017uul}, which is relevant in photonics.
In addition, our proposal does not consider noninertial frame under rotation, but the effective metric is generated by materials. 
It may be tested in magnetoelectric materials such as LiCoPO$_4$, TbPO$_4$~\cite{0022-3727-38-8-R01,Rivera2009}, and ZnCr$_2$Se$_4$ spinel with a conical spiral state~\cite{0953-8984-20-43-434203}.
``Magnetic fields" can experimentally be generated at a surface of magenetoelectirc materials, 
or by preparing inhomogeneous ordering such as domain walls~\cite{PhysRevLett.95.237402}.

\section{geodesic equation of light}
We discuss the propagation of a monochromatic electromagnetic wave with frequency $\omega$ in anisotropic, inhomogeneous, and lossless mediums exhibiting linear magnetoelectric effect in Gaussian units:
\bea
4\pi\bm P &=&  \chi_e \bm E-\bm g\times\bm H ,
\label{eq:polarization}\\
4\pi\bm M &=&\chi_\mu \bm H+\bm g\times\bm E ,
\label{eq:magnetization}
\eea
where $\bm P$ and $\bm M$ ($\bm E$ and $\bm H$) are polarization, and magnetization (electric and magnetic fields), respectively.
$\chi_e$, $\chi_\mu$, and $\bm g$ represent electric susceptibility, magnetic susceptibility, and  a vector component of magnetoelectric tensors~\cite{0022-3727-38-8-R01,0953-8984-20-43-434203}.
We first assume that $\bm g$ slowly varies on space, and show that $\nabla_x\times\bm g$ plays a role of ``magnetic field" in the geodesic equation of a light.

By taking the polarization and magnetization in Eqs.~\eqref{eq:polarization} and~\eqref{eq:magnetization} into account, the Maxwell equations read
\bea
\nabla_x\cdot\left(\epsilon \bm E-\bm g\times\bm H \right)&=&0 ,\\
\nabla_x\cdot\left(\mu \bm H +\bm g\times\bm E \right)&=&0 ,\\
\nabla_x\times \bm E+\frac{1}{c}\partial_t\left(\mu \bm H +\bm g\times\bm E \right)&=& 0 ,\\
\nabla_x\times \bm H-\frac{1}{c}\partial_t\left(\epsilon \bm E-\bm g\times\bm H \right) &=& 0,
\eea
where $\epsilon$ and $\mu$ are the dielectric permittivity and magnetic permeability, and $c$ is the speed of light in vacuum.
Those equations become the same as the Maxwell equations in rotating frame~\cite{Landau} if $\bm g=\bm\Omega\times\bm x/2$ with angular velocity $\bm\Omega$, and the centrifugal force terms [${\cal O}(\bm\Omega^2$) terms] are neglected.
For the monochromatic wave electric and magnetic fields, $\bm E=\tilde{\bm E}(\omega)e^{-i\omega t}$ and $\bm H=\tilde{\bm H}(\omega)e^{-i\omega t}$, we obtain the eigen equations:
\bea
\bm D\times \mu^{-1}\left(\bm D\times \tilde{\bm E}\right)-\epsilon \tilde{\bm E} &=&0 , 
\label{eq:electric} \\
\bm D\times \epsilon^{-1}\left(\bm D\times \tilde{\bm H}\right)-\mu \tilde{\bm H} &=&0 , 
\label{eq:magnetic} 
\eea
where $\bm D=\lambdabar\nabla_x-i \bm g$, and $\lambdabar=\lambda/(2\pi)=c/\omega$. 
Because of the inhomogeneous magnetoelectric effect, $\bm D$ no longer commute with each other and satisfy
\bea
[D_i,D_j]=-i \lambdabar \epsilon_{ijk}\bm b_k,
\eea
where $\bm b=\nabla_x\times \bm g$, and $\epsilon_{ijk}$ is the totally antisymmetric tensor.
The noncommutativity leads to the Coriolis force in the dynamics of electromagnetic waves. 

We consider the eikonal approximation in the linear order of $\lambdabar$, and derive the geodesic equation of a light with the Berry curvature correction.
When $\bm g=0$, and the anisotropy and inhomogeneity of $\epsilon$ and $\mu$ are small and  treated perturbatively, 
as performed e.g., in Refs.~\cite{PhysRevB.72.035108,PhysRevA.75.053821,Bliokh},
by introducing dimensionless momentum operator $\hat{\bm q}=-i\lambdabar\nabla_x$ and rewriting Eq.~\eqref{eq:electric} into the Sch{\"o}dinger-type equation ${\cal H} \tilde{\bm E}=0$,
we can diagonalize the $3\times3$ matrix ${\cal H}$, and obtain three eigenvectors. 
Two of them correspond to transverse modes, and the other corresponds to the longitudinal (resonant) mode.  
Then when the system is off-resonant ($\epsilon\neq0$)~\cite{PhysRevA.75.053821}, by neglecting the resonant mode, we can introduce the Berry connection $\bm\Lambda$ and Berry curvature $\bm\Omega$ to describe the noncommutative dynamics in the projected space spanned  by the transverse modes in the same way with quantum mechanics: 
\bea
[\hat{\bm x}_i,\hat{\bm x}_j]=i\lambdabar \epsilon_{ijk}\bm\Omega_k ,
\eea
where $\hat{\bm x}=i\lambdabar\nabla_q+\bm \Lambda$ are covariant coordinate operators~\cite{PhysRevB.72.035108,PhysRevA.75.053821}.
In the geodesic equation, the noncommutavity is implemented as the spin-orbit coupling~\cite{PhysRevB.72.035108,PhysRevA.75.053821,Bliokh}.

Now we take the effect of $\bm g$ into account based on the derivative expansion. 
This can be done in the same way with the vector potential in the wave packet dynamics in electron systems~\cite{PhysRevB.59.14915,PhysRevB.72.085110,PhysRevB.77.035110,PhysRevB.95.125137}.
The modification is straightforward: We replace the canonical momentum $\bm q=-i\lambdabar\nabla_x$ by the covariant momentum $\bm p=-i\bm D$ as $\bm q=\bm p+\bm g$ in the effective Lagrangian given in Ref.~\cite{Bliokh}.
For simplicity, we hereafter consider a locally isotropic medium in which the dynamics becomes Abelian~\cite{PhysRevLett.93.083901,PhysRevE.74.066610,PhysRevB.72.035108,PhysRevA.75.053821,Bliokh}.
The geodesic equation reads, in the linear order of $\lambdabar$ and $\bm g$,
\bea
\dot{\bm x}_c&=&\frac{\bm p_c}{p_c} -\dot{\bm p}_c\times \langle \eta_c |\lambdabar\bm \Omega |\eta_c\rangle,
\label{eq:eom_x} \\
\dot{\bm p}_c&=&\bm e +\dot{\bm x}_c\times \bm b ,
\label{eq:eom_p} \\
|\dot{\eta}_c\rangle&=&i \dot{\bm p}_c\cdot \bm \Lambda |\eta_c\rangle ,
\label{eq:eom_hel}
\eea
where $\bm x_c$ and $\bm p_c$ are the coordinates and dimensionless wave vectors of a ray of light.
$|\eta_c\rangle=(\eta_+,\eta_-)^t$ represents polarization states of lights.
We use right-handed (+) and left-handed (-) circularly polarized waves as a basis of polarization.
The dot means the derivative with respect to the ray length $l$, namely, the derivative along the trajectory, not time~\cite{PhysRevB.72.035108,PhysRevA.75.053821,Bliokh}.
$n(\bm x_c)$ is isotropic and slowly varying refractive index, and $\bm e=\nabla_{x_c}n(\bm x_c)$. 
$\bm\Lambda$ and $\bm\Omega=\nabla_{\bm p_c}\times\bm\Lambda-i\bm\Lambda\times\bm\Lambda$ are the Berry connection and Berry curvature, and given explicitly as~\cite{PhysRevLett.93.083901,PhysRevE.74.066610,PhysRevB.72.035108,PhysRevA.75.053821,Bliokh}
\bea
\bm\Lambda=\frac{1}{p \tan\theta}\bm e_\vphi\sigma_z ,\;\;\;\;
\bm\Omega=-\frac{\bm p}{p^3}\sigma_z ,
\label{eq:berry}
\eea
where $p$, $\theta$ and $\vphi$ are spherical coordinates in $\bm p$ space, and $\sigma_z$ is the Pauli matrix.
In the above geodesic equation, $\bm b$ is the same as the magnetic field in the classical equation of motion of electrons, so that we term $\bm b$ ``magnetic field."
However it originates from the effective metric induced by magnetoelectric materials, and strictly speaking, is equal to gravitomagnetic field~\cite{Hidaka}.
The same is true for ``electric field" $\bm e=\nabla_{x_c}n(\bm x_c)$, and it is equal to gravitoelectric field~\cite{Hidaka}.

We note that we neglected a self-rotation of a wave packet and the associated ``Zeeman energy," 
which may not be negligible in a beam with intrinsic angular momentum such as optical vortices.
We also note that the same idea to use the optical magnetoelectric effect as ``vector potential" of a light has been indicated in Ref.~\cite{PhysRevLett.95.237402}
(For the explicit relation to ours, see~\footnote{See Supplemental Material at [URL will be inserted by publisher] for the derivation of the optical magnetoelectric effect}).
However, they employed the conventional Fermat's variational principle with optical Lorentz force, 
and topological phenomena caused by ``magnetic fields" were not discussed. 
Such topological phenomena including the chiral magnetic effect can be analyzed only by taking the effect of the Berry curvature of photons into account, and are the main subject of this paper.
Moreover, we will analyze the chiral magnetic effect of a light beyond the geometrical optics approximation.

\section{Chiral magnetic effect of light}  
We discuss anomalous shift of a wave packet of light caused by the interplay between ``magnetic fields" and Berry curvature.
The geodesic equation in Eqs.~\eqref{eq:eom_x} and~\eqref{eq:eom_p} is diagonal for right- and left-handed polarizations, and we obtain
\bea
\sqrt{J}\dot{\bm x}_\chi&=& \frac{\bm p_\chi}{p_\chi}+\chi\lambdabar\bm \Omega_{++}\times \bm e-\chi\bm b\left(\lambdabar\bm \Omega_{++}\cdot \frac{\bm p_\chi}{p_\chi}\right) ,
\label{eq:sol_x}\\
\sqrt{J}\dot{\bm p}_\chi&=&
\bm e- \bm b\times  \frac{\bm p_\chi}{p_\chi}-\chi\lambdabar \bm \Omega_{++}\left(\bm e\cdot \bm b\right)  ,
\label{eq:sol_p} 
\eea
where $\sqrt{J}=1-\chi\bm b\cdot \lambdabar\bm \Omega_{++}$, and $\chi=\pm 1$ for right-handed (left-handed) polarization.
The second term in Eq.~\eqref{eq:sol_x} leads to the optical Hall effect~\cite{PhysRevLett.93.083901,PhysRevE.74.066610,PhysRevA.45.8204,PhysRevA.46.5199}. 
The last terms in Eqs.~\eqref{eq:sol_x} and~\eqref{eq:sol_p} lead to the analog of the chiral magnetic effect~\cite{PhysRevLett.109.162001,PhysRevLett.109.181602,PhysRevLett.110.262301,PhysRevB.88.104412} and the spectral flow~\cite{PhysRevB.88.104412,NIELSEN1983389}. 

Let us first discuss the chiral magnetic effect of a light.
We consider the propagation of a wave packet of a light. 
As in Eq.~\eqref{eq:sol_x}, the anomalous group velocity parallel to $\bm b$ is generated by the Berry curvature.
Since the sign is opposite between the right- and left-handed polarizations, those states propagate along the opposite direction parallel to $\bm b$. 
We term the helicity-dependent shift of a wave packet chiral magnetic effect of a light since $\bm b$ in the geodesic equation is the same as the magnetic field in the classical equation of motion of electrons~\cite{PhysRevB.88.104412}. 

\begin{figure}[t]
 \includegraphics[width=.42\textwidth]{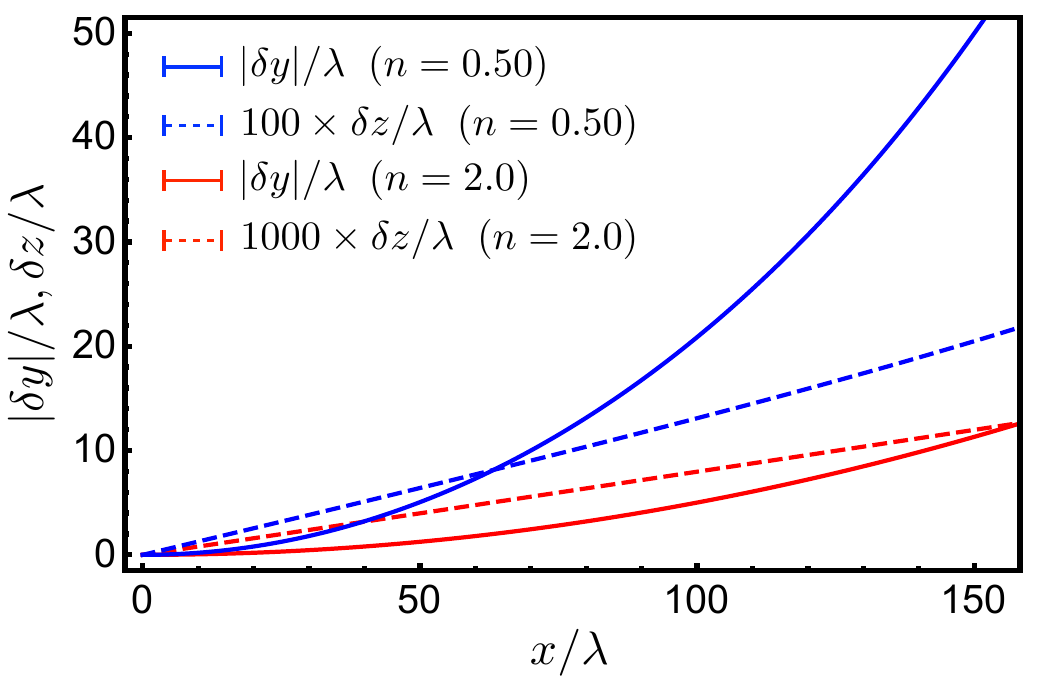}
 \caption{
Shift of the center of wave packets by ``Lorentz force" and chiral magnetic effect of a light.
The blue (red) solid and dashed lines are shifts in the $y$ and $z$ directions at $n=2.0$ ($0.50$) and $\bm b\lambda =(0,0, 2\times 10^{-3})$, respectively.
\label{fig1}
}
\end{figure}
To confirm the idea, we consider a physical set up similar to Ref.~\cite{PhysRevLett.95.237402}, and numerically calculate the trajectories of wavepackets propagating through thin films at $\bm b\lambda=(0,0, 2\times 10^{-3})$~\footnote{We estimate the gradient $b=(\nabla\times\bm t)_z$ as $b\lambda\sim\alpha\lambda/L\sim2\times 10^{-3}$, where $\alpha=0.2$, $L=50\mu$m, and $\lambda=500$nm are typical magnitude of the magnetoelectric tensors~\cite{Rivera2009}, thickness of a sample, and wavelength of an incident light, respectively.} and uniform $n=0.50$, $2.0$ ($\bm e=0$). 
We assume that the sample is infinitely large in the $yz$ plane and thin along the $x$ direction. 
Then we consider the incident light along the $x$ direction with right-handed polarization. 
Without $\bm b$ field, the light propagates along the $x$ direction.
In the presence of $\bm b$ field, shifts along the $y$ and $z$ directions occur because of the ``Lorentz force," and chiral magnetic effect of a light. 
The numerical results are shown in Fig.~\ref{fig1}.
The quadratic curves due to the Lorentz force are independent of the polarization, and reproduce the result of Ref.~\cite{PhysRevLett.95.237402}. 
In addition, we find the linear and helical shift parallel to $\bm b$.
This is the consequence of the chiral magnetic effect of a light, and can be analyzed only by taking the Berry curvature correction into account.
From the slope, we estimate the shift of the wave packet as $\delta z=\pm0.05\mu$m for right-handed (left-handed) polarization when 
the sample thickness is $x=50\mu$m and $n=0.50$.

\begin{figure}[t]
\centering
 \includegraphics[width=.33\textwidth]{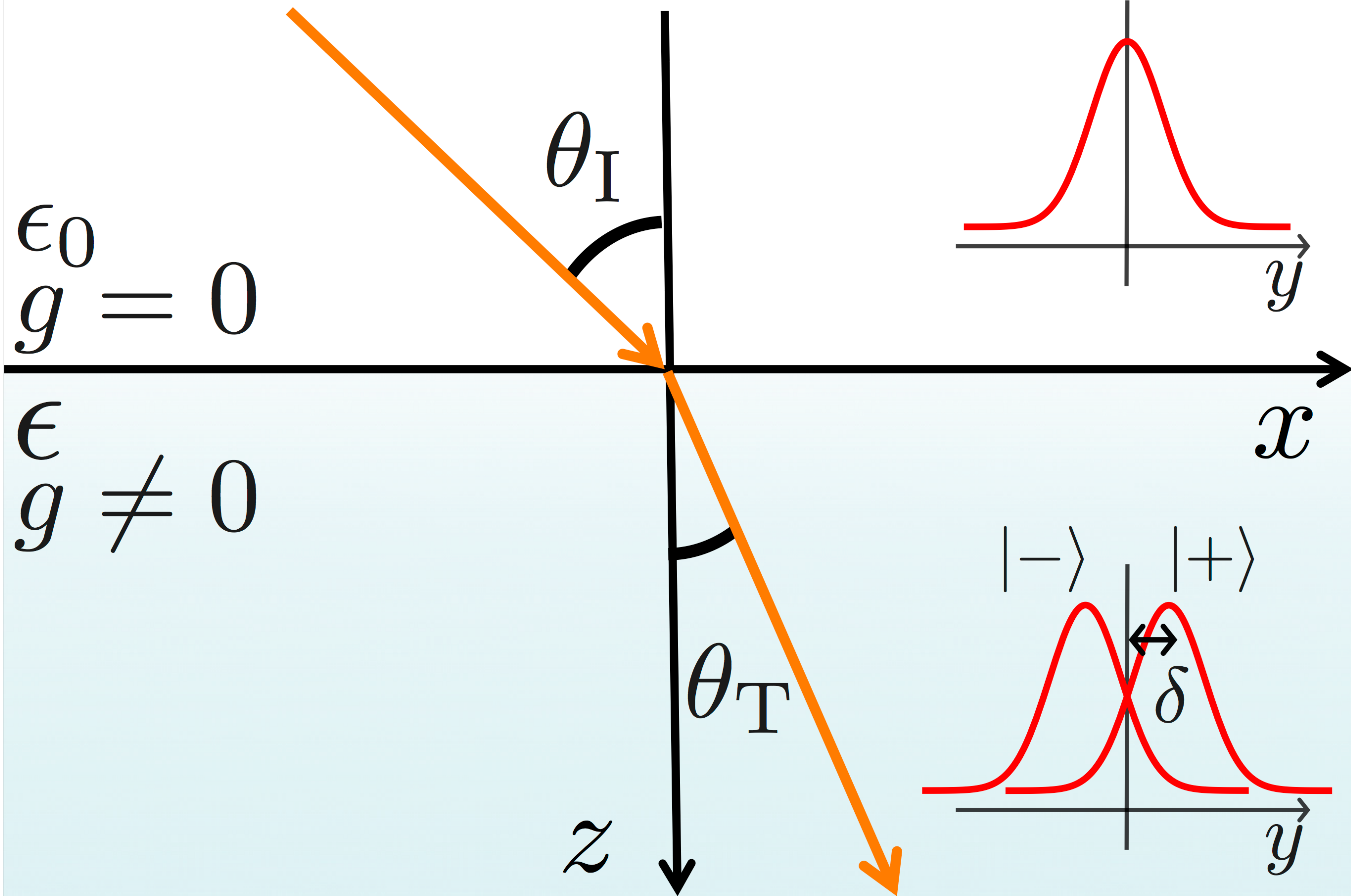}
\caption{
Schematic figure of the incident, and refracted lights.
Center position of a wave packet is displaced upon transmission via spin-orbit coupling.
\label{fig2}
}
\centering
 \includegraphics[width=.45\textwidth]{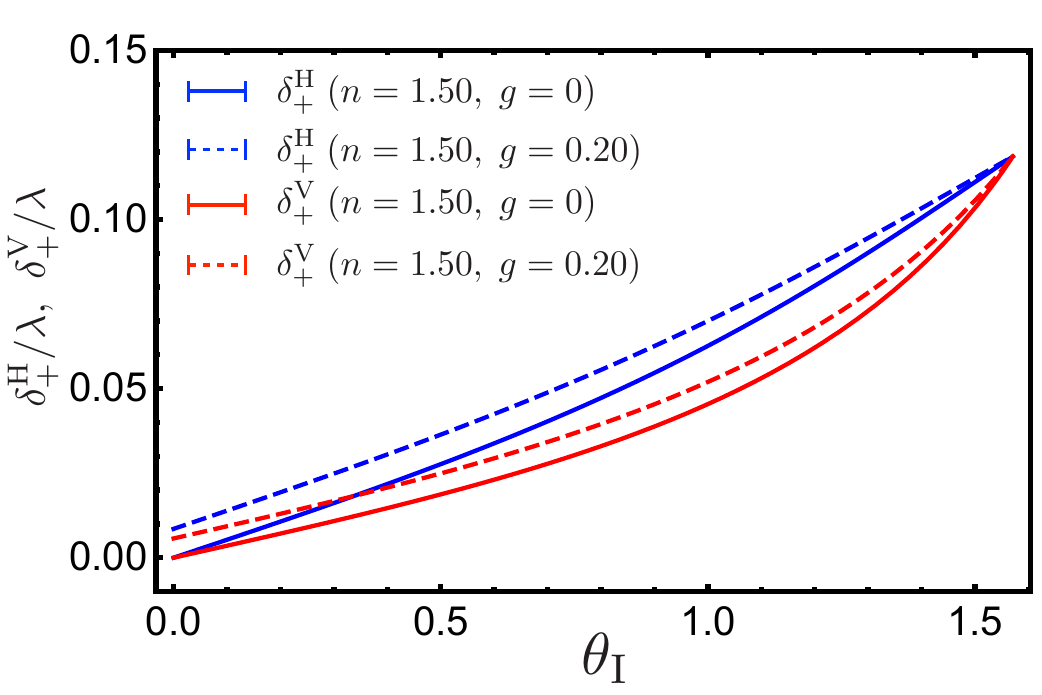}
\caption{
Incident angle dependence of displacements $\delta^{\rm H}_+$ and $\delta^{\rm V}_+$
at $n=1.50$, and $g=0$, $0.20$.
The solid (dashed) curve shows the displacements at $g=0$ ($g=0.20$).
\label{fig3}
}
\end{figure}

For quantitative prediction to experiments, we consider a realistic physical set up based on an experimental work~\cite{Hosten787}, and compute the transverse displacements of a wave packet refracted at a surface between vacuum and a uniform magnetoelectric material.
We consider a wave packet incident with angle $\theta_{\rm I}$ to the $z=0$ surface  between vacuum ($z<0$) and a half-infinite magnetoelectric material ($z>0$) with $\bm g=(g,0,0)$ [$g>0$] (see Fig.~\ref{fig2}). 
At a sharp surface, $n$ and $\bm g$ rapidly change like the step function, and the delta function-like ``electric field" $\bm e=\nabla n\sim \delta(z)\hat{z}$ and ``magnetic field" $\bm b=\nabla_x\times \bm g\sim\delta(z)\hat{y}$ are induced at a surface.
Then, from the analysis based on the Berry curvature, we expect shifts of a wave packet due to the spin Hall/chiral magnetic effect of a light.
However, when $n$ and $\bm g$ rapidly change, we no longer employ the geometrical optics approximation and need to directly solve the Maxwell equations. 
As will be shown, the shift parallel to $\bm b$ still arises, and the above analysis based on the Berry curvature is qualitatively correct. 
Following Ref.~\cite{Hosten787}, we consider a wavepacket with finite distribution along the $y$ direction.
For a horizontally-polarized wave, the amplitude reads~\cite{Hosten787}
\bea
\Psi_{\rm initial}=\frac{\Psi(y)}{\sqrt{2}}
\begin{pmatrix}
1 \\
1
\end{pmatrix}
=
\int dp_y e^{ip_yy}\frac{\Phi(p_y)}{\sqrt{2}}
\begin{pmatrix}
1 \\
1
\end{pmatrix} .
\label{eq:initial}
\eea
The right- and left-handed states rotate in an opposite way upon transmission to satisfy the transversality condition, which leads to the transverse displacement of a central position via the spin-orbit coupling~\cite{Hosten787} as schematically shown in Fig.~\ref{fig2}:
\bea
\Psi_{\rm final}&=&t_p\int dp_y e^{ip_yy}\frac{\Phi(p_y)}{\sqrt{2}}
\begin{pmatrix}
e^{-i p_y\delta^{\rm H}} \\
e^{+i p_y\delta^{\rm H}}
\end{pmatrix}
\notag  \\
&=&\frac{t_p}{\sqrt{2}}
\begin{pmatrix}
\Psi(y-\delta^{\rm H}) \\
\Psi(y+\delta^{\rm H}) 
\end{pmatrix}, 
\label{eq:final}
\eea
where $t_p$ is the Fresnel coefficient. 
Similar arguments hold for a vertical polarization and lead to the transverse displacement shown below.
By solving the transmission problem with the polarization in Eq.~\eqref{eq:polarization} and magnetization in Eq.~\eqref{eq:magnetization}, the displacements of the spin components ($\pm$) for horizontally and vertically polarized waves are given as 
\bea
\delta^{\rm H}_\pm&=&\pm\frac{\lambda}{2\pi}\frac{\cos\tilde{\theta}_{\rm T}-(t_s/t_p)\cos\tilde{\theta}_{\rm I}}{\sin\tilde{\theta}_{\rm I}} ,
\label{eq:displacement1}
\\
\delta^{\rm V}_\pm&=&\pm\frac{\lambda}{2\pi}\frac{\cos\tilde{\theta}_{\rm T}-(t_p/t_s)\cos\tilde{\theta}_{\rm I}}{\sin\tilde{\theta}_{\rm I}},
\label{eq:displacement2}
\eea
where $\tilde{\theta}_{\rm I}=\theta_{\rm I}+g\cos\theta_{\rm I}/(n^2-1)$, and $\lambda$, $\theta_{\rm I}$, $\tilde{\theta}_{\rm T}$, $t_s$, and $t_p$ are the wave length of a light in incident medium, the incident and refraction angles, and the Fresnel coefficients, respectively~\footnote{See Supplemental Material at [URL will be inserted by publisher] for computational details}.
$\tilde{\theta}_{\rm T}$ is related with $\theta_{\rm I}$ by Snell's law as $\sin\tilde{\theta}_{\rm T} =\sin\tilde{\theta}_{\rm I}/n$.
The chiral magnetic effect can be computed as the spin Hall effect in the boosted frame, so that  
the transverse displacements in Eqs.~\eqref{eq:displacement1} and~\eqref{eq:displacement2} have the same form with Ref.~\cite{Hosten787} except that the incident angle is modified by the magnetoelectric effect.

We show the displacements of the right-handed component ($+$) at zero and nonzero $g$ in Fig.~\ref{fig3}. 
Wave packets experience the displacements due to the chiral magnetic effect only at nonzero $g$.
We find that the key signal of the chiral magnetic effect is the nonvanishing of $\delta^{\rm H,V}_\pm$ at $\theta_{\rm I}=0$, namely, when the beam is normally incident to the material. 
This can be qualitatively understood from the equation of motion~\eqref{eq:sol_x}.
The spin Hall effect is originated from $\chi\bm \Omega_{++}\times\bm e\sim-\chi\hat{\bm p}\times\hat{z}$, 
and vanishes when $\bm p$ is normal to a surface.
On the other hand, the chiral magnetic effect is originated from $-\chi\bm b(\bm \Omega_{++}\cdot \bm p)\sim \chi\hat{y}$,
and is approximately independent of the direction of $\bm p$.
When $\lambda=633$nm~\cite{Hosten787}, the intercepts are estimated as $\delta^{\rm H}_+(\theta_{\rm I}=0)=5.4$nm, and $\delta^{\rm V}_+(\theta_{\rm I}=0)=3.6$nm, which are observable shifts via quantum weak measurements~\cite{Hosten787}. 

Finally, we discuss a photonic analog of the spectral flow, which is also referred to as the Adler-Jackiw anomaly.
In chiral fermions, when the pseudosclar product of electromagnetic fields $\bm E\cdot \bm B$ is nonzero, the excitation from the left-handed fermions to right-handed fermions occurs and the chirality imbalance is dynamically generated~\cite{PhysRevB.88.104412,NIELSEN1983389}.
The phenomena has been utilized an experimental signal of Weyl or Dirac points in transport phenomena in condensed matter materials~\cite{PhysRevB.88.104412,PhysRevX.5.031023}.
We expect an analogous effect when $\bm e\cdot\bm b\neq0$.
As in Eq.~\eqref{eq:sol_p}, $\bm e\cdot\bm b$ appears in $\dot{\bm p}$, so that 
lights are accelerated/decelated depending on their helicities. 
From Eq.~\eqref{eq:eom_hel}, the integration of $\dot{\bm p}\cdot\bm\Lambda$ over the trajectories,
\bea
\Theta_\pm=\int dl \frac{d\bm p_{\pm}}{dl}\cdot [\bm\Lambda]_{\pm\pm}=\int d\bm p_{\pm}\cdot [\bm\Lambda]_{\pm\pm} ,
\eea
 contributes to the phase of the outstate as $|\eta_c^{\rm out}\rangle=[e^{i\Theta_+}\eta_+^{\rm in},e^{i\Theta_-}\eta_-^{\rm in}]^t$,
with the initial polarization $|\eta_c^{\rm in}\rangle=[\eta_+^{\rm in},\eta_-^{\rm in}]^t$.
Therefore, when $\bm e\cdot\bm b\neq0$, phase difference arises ($\Theta_+\neq\Theta_-$) 
because of the aforementioned helicity-dependent acceleration/deceleration.
This is one approach to discuss a photonic analog of the spectral flow.
A similar phase shift due to inhomogeneous magnetoelectric effect has been discussed in topological insulators~\cite{PhysRevB.78.195424,PhysRevLett.105.166803,PhysRevLett.105.057401,PhysRevLett.108.087403,Okada}, which involve the scalar component of magnetoelectric tensors~\cite{0022-3727-38-8-R01,0953-8984-20-43-434203}.

\section{Summary} 
We have studied the photonic analog of the chiral magnetic (vortical) effect. 
We discuss that rotation of the vector component of magnetoelectric tensors $\bm b=\nabla\times\bm g$ behaves as ``magnetic field" of a light~\cite{PhysRevLett.95.237402}. 
The interplay between ``magnetic fields" and Berry curvature of photons causes helical shifts of a wave packet along the direction parallel to  ``magnetic fields."
This is the chiral magnetic effect of a light in geometric optics. 
We have confirmed the chiral magnetic effect of a light arises even when geometric optics breaks down, by directly solving the transmission problem of a wave packet at a surface of a magnetoelectric material. 
We show that the signal of the chiral magnetic effect is the nonvanishing of displacements for the beam normally incident to the material.

There are several generalizations of our paper.
One direction is photonic crystals with artificaial magnetic fields~\cite{PhysRevLett.100.013904,PhysRevA.78.033834,PhysRevLett.100.013905,Hafezi2011,Fang2012,Hafezi2013,Lu2014}, in which the chiral magnetic effect of a light might be enhanced like the optical Hall effect~\cite{PhysRevLett.93.083901,PhysRevE.74.066610}.
Another direction is a generalization to include another form of the Berry curvature, which involves temporal derivative, and is understood as emergent electric fields in momentum space~\cite{RevModPhys.82.1959}.
It causes anomalous transport effects such as the Thouless pumping~\cite{PhysRevB.27.6083}.
It is interesting to discuss a photonic analog of the Thouless pumping, which may give a sizable effect due to the massless nature of photons, as in the case of Weyl semimetals~\cite{PhysRevLett.117.216601}.


\begin{acknowledgements}
The author thanks K.~Fukushima, Y.~Hidaka, and S.~Nakamura for useful comments.
This work was supported by JSPS Grant-in-Aid for Scientific Research (No: JP16J02240).
\end{acknowledgements}

\bibliography{./photon}

\end{document}


\title{Chiral magnetic effect of light: Supplemental material}

\author{Tomoya Hayata}
\affiliation{
Department of Physics, Chuo University, 1-13-27 Kasuga, Bunkyo, Tokyo, 112-8551, Japan 
}


\maketitle

\section{Nonreciprocal refractive index}
\label{sec:refractive} 
We here clarify the relation between our proposal of the ``magnetic fields" of a light and the one given in Ref.~\cite{PhysRevLett.95.237402}.
For this purpose, we derive the dispersion relations of electromagnetic waves satisfying the Maxwell equations in the presence of magnetoelectric effects, whose explicit forms are given in Eqs.~($3$)-($6$) in the main text.
We consider the case that the electric susceptibility $\epsilon$ and magnetic susceptibility $\mu$ are isotropic and spatially uniform, and the vector component of magnetoelectric tensors $\bm g$ is spatially uniform.
Then the Maxwell equations~($11$) and~($12$) in the main text are written as
\bea
\left(-\bm D^2-\epsilon\mu \right)\bm E &=&0 , 
\label{eq:electric} \\
\left(-\bm D^2-\epsilon\mu \right)\bm H &=&0 , 
\label{eq:magnetic} 
\eea
where $\bm E=\tilde{\bm E}e^{i\bm q\cdot\bm x-i\omega t}$ and $\bm H=\tilde{\bm H}e^{i\bm q\cdot\bm x-i\omega t}$,
and $\bm D=i(c/\omega)\bm q-i \bm g$. 
The on-shell condition $\bm D^2+\epsilon\mu=0$ is the quadratic equation with respect to $\omega$, and solved as
\bea
\omega_\pm=(c^\prime)^2\left(-\frac{1}{c}\bm g\cdot \bm q\pm\sqrt{\frac{\bm q^2}{(c^\prime)^2}+\left(\frac{1}{c}\bm g\cdot \bm q\right)^2}\right) ,
\label{eq:dispersion}
\eea
where $c^\prime=c/\sqrt{\epsilon\mu-\bm g^2}$. 
The dispersion relations are nonreciprocal, namely, depend on the propagating direction if $\bm g\cdot\bm q\neq0$.
We study Eq.~\eqref{eq:dispersion} in detail for two cases: (I) $|\bm g|\sim\sqrt{\epsilon\mu}$ ($c^\prime/c\gg1$). In this case, we have
\bea
\omega_+&=&\frac{c}{2g}q_x+\frac{c}{2g q_x}(q_y^2+q_z^2) ,
\\
\omega_-&=&-2\frac{(c^\prime)^2}{c}g q_x-\frac{c}{2g q_x}(q_y^2+q_z^2) ,
\label{eq:dispersion2}
\eea
where we assumed $\bm g=(g,0,0)$, and $g q_x>0$. 
Photon shows quadratic (nonrelativistic) dispersion relations along the direction perpendicular to $\bm g$, while it shows the linear dispersion relations along the direction parallel to $\bm g$.
Next let us consider the case: (II) $|\bm g|\ll\sqrt{\epsilon\mu}$. In this case, we have
\bea
\omega_\pm=\pm \bar{c} |\bm q|\left(1\mp\frac{\bar{c}}{c}\bm g\cdot \hat{\bm q}\right) ,
\label{eq:dispersion3}
\eea
where $\hat{\bm q}=\bm q/|\bm q|$, and $\bar{c}= c/\sqrt{\epsilon\mu}$. 
The effective refractive index $n_\pm$ is defined as $\omega_\pm=\pm (c/n_\pm)|\bm q|$, and we get
\bea
n_+-n_-=2\bm g\cdot \hat{\bm q} .
\label{eq:refractive}
\eea
This is nothing but the optical magnetoelectric effect in Ref.~\cite{PhysRevLett.95.237402}, 
and the origin of the ``Lorentz force" in their proposal.
Therefore our proposal to use the inhomogeneous magnetoelectric effect is the same as the one in Ref.~\cite{PhysRevLett.95.237402}.

\section{Chiral magnetic effect at an intersurface}  
\begin{figure}[!t]
 \includegraphics[width=.48\textwidth]{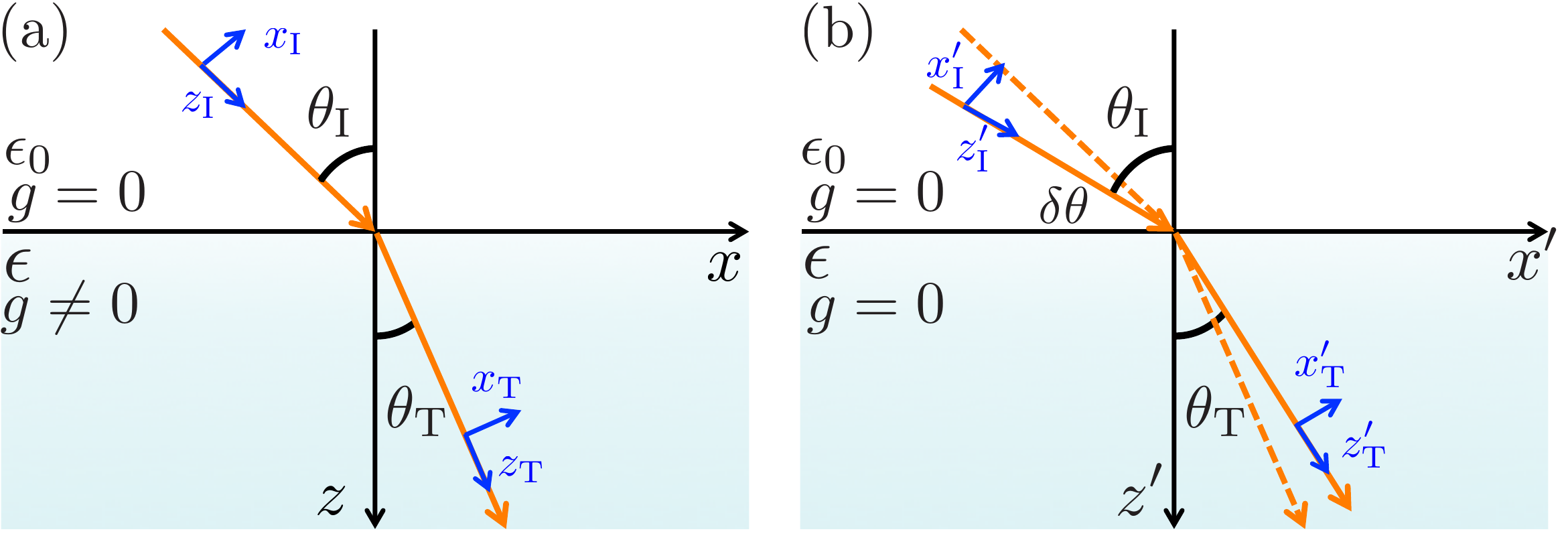}
\caption{
Schematic figures of the incident, and refracted lights in the laboratory frame (a) and ``rest" frame (b).
Blue arrows denote the coordinates defined by the central wave vectors.
\label{fig1}
}
\end{figure}
We here compute transverse displacements of a wave packet in an experimental setup discussed in Ref.~\cite{Hosten787}, with taking the magnetoelectric effect into account. 
We consider the transmission of a monochromatic wave packet through the $z=0$ boundary from vacuum ($z<0$) to a half-infinite magnetoelectric medium ($z>0$) [See Fig.~\ref{fig1}(a)].
At a sharp surface between them, $n=\sqrt{\epsilon\mu}$ and $\bm g$ rapidly change like the step-function, and the delta-function-like effective electric field $\bm e=\nabla n\sim \delta(z)\hat{z}$ and magnetic field $\bm b=\nabla\times \bm g\sim\delta(z)\hat{y}$ are generated  (Hereafter we consider $\bm g=(g,0,0)$ [$g>0$]). 
Then from the analysis based on geometric optics with the Berry curvature correction, 
we expect anomalous shifts of a wave packet by the spin Hall/chiral magnetic effect of a light.
However in a situation where $n$ and $\bm g$ rapidly changes, the geometrical optics approximation breaks down, and we need to solve the transmission problem on the basis of the Maxwell equations.

The Maxwell equations~($3$)-($6$) in the main text have the same form with those in moving dielectrics~\cite{Landau2} with the velocity $\bm\beta=\bm v/c=-\bm g/(n^2-1)$ when ${\rm O}(\bm \beta^2)$ terms are negligible,
so that transverse displacements can easily be computed by considering the spin Hall effect in the ``rest" frame, which moves with the surface of a virtual dielectric material.
Following Ref.~\cite{Hosten787}, we consider a polarized incident beam with finite distribution along the $y$ direction. 
The wave vector of the incident beam is represented as $\bm q_{\rm I}=q_{\rm I}(\hat{z}_{\rm I}+\nu_{{\rm I}y} \hat{y})$, where $\nu_{{\rm I}y}=q_{{\rm I}y}/q_{\rm I}$ ($q_{{\rm I}y}/q_{\rm I}\ll1$), and $q_{{\rm I}y}$ has a distribution around zero.
As a model of polarization of the incident beam, we employ~\cite{Hosten787,PhysRevE.75.066609}
\bea
\tilde{\bm E}_{\rm I}=\frac{E_0}{\sqrt{1+|m|^2}}\left(\hat{x}_{\rm I}+m\hat{y}-m\nu_{{\rm I}y}\hat{z}_{\rm I}\right) ,
\label{eq:incident}
\eea
where $\hat{x}_{\rm I}$ is the unit vector along the $x_{\rm I}$ direction, and $(x_{\rm I},y,z_{\rm I})$ are the coordinates defined by using the central wave vector $\bm q_{{\rm I}c}=q_{\rm I}\hat{z}_{\rm I}$ of the incident beam [See Fig.~\ref{fig1}(a)].
We can change the frame by considering the Lorentz transformation with the boost parameter $-\bm\beta=-\beta\hat{x}=-\beta\cos\theta_{\rm I}\hat{x}_{\rm I}-\beta\sin\theta_{\rm I}\hat{z}_{\rm I}$ [$\beta=g/(n^2-1)>0$].
Then the incident electric field in the ``rest" frame is given as
$\tilde{\bm{{\cal E}}}_{\rm I}=\tilde{\bm E}_{\rm I}-\bm\beta\times \tilde{\bm H}_{\rm I}$, with $\tilde{\bm H}_{\rm I}=c\bm q_{\rm I}\times\tilde{\bm E}_{\rm I}/\omega$:
\bea
\begin{split}
\tilde{\bm{{\cal E}}}_{\rm I}&=\frac{{\cal E}_0}{\sqrt{1+|m|^2}}\Bigl(\left(1+m\beta\cos\theta_{\rm I}\nu_{{\rm I}y}^\prime\right)\hat{x}^\prime_{\rm I}
\\
&+\left(m-\beta\cos\theta_{\rm I}\nu_{{\rm I}y}^\prime\right)\hat{y}
-m\nu_{{\rm I}y}^\prime\hat{z}_{\rm I}^\prime\Bigr) ,
\end{split}
\label{eq:incident_boost}
\eea
where ${\cal E}_0=(1+c\bm\beta\cdot\bm q_{\rm I}/\omega)E_0$, $\nu_{{\rm I}y}^\prime=(1-\beta\sin\theta_{\rm I})\nu_{{\rm I}y}$, 
and we neglected ${\cal O}(\beta^2)$ and ${\cal O}(\nu_{{\rm I}y}^2)$ terms.
The new coordinates $(x^\prime_{\rm I},y,z^\prime_{\rm I})$ are tilted by the angle $\delta\theta=\beta\cos\theta_{\rm I}$ from the co-moving coordinates $(\bar{x}_{\rm I},\bar{y},\bar{z}_{\rm I})$  [See Fig.~\ref{fig1}(b)]: 
\bea
\hat{x}^\prime_{\rm I} &=& \hat{\bar{x}}_{\rm I}-\beta\cos\theta_{\rm I}\hat{\bar{z}}_{\rm I},
\\
\hat{z}^\prime_{\rm I} &=& \hat{\bar{z}}_{\rm I}+\beta\cos\theta_{\rm I}\hat{\bar{x}}_{\rm I},
\eea
where we neglected ${\cal O}(\beta^2)$ terms.
Because of the transverse Lorentz boost (the Lorentz transformation along the $x_{\rm I}$ direction),
the momentum is shifted along the $x_{\rm I}$ direction:
$\bm q^\prime_{\rm I}=q^\prime_{\rm I}\left(\hat{z}_{\rm I}^\prime+\nu_{{\rm I}y}^\prime \hat{y}\right)$,
where $q^\prime_{\rm I}=(1+\beta\sin\theta_{\rm I})q_{\rm I}$, so that the polarization is also rotated to satisfy the transversality condition $\bm q^\prime_{\rm I}\cdot \tilde{\bm{{\cal E}}}_{\rm I}=0$ as in Eq.~\eqref{eq:incident_boost}.
In the boosted frame, the electromagnetic field with the polarization~\eqref{eq:incident_boost} is incident upon a virtual dielectric medium with the modified angle $\tilde{\theta}_{\rm I}=\theta_{\rm I}+\delta\theta$  [See Fig.~\ref{fig1}(b)].

Now we discuss the transmission of a light.
The eigenstates of transmission are $s$- and $p$-polarization states defined by using the local beam coordinates of $\bm q_{\rm I}^\prime$ [$(x^{\prime\prime}_{\rm I},y^{\prime\prime}_{\rm I},z^{\prime\prime}_{\rm I})$], not those of the central wave vector $\bm q_{{\rm I}c}^\prime$~\cite{Hosten787,PhysRevE.75.066609}.
The central wave vector coordinates $(x^\prime_{\rm I},y,z^\prime_{\rm I})$ is related with the local beam coordinates $(x^{\prime\prime}_{\rm I},y^{\prime\prime}_{\rm I},z^{\prime\prime}_{\rm I})$ by the following relations:~\cite{Hosten787,PhysRevE.75.066609}
\bea
\hat{x}^{\prime\prime}_{\rm I} &=& \hat{x}^\prime_{\rm I}+\nu_{{\rm I}y}^\prime\cot\tilde{\theta}_{\rm I}\hat{y},
\\
\hat{y}^{\prime\prime}_{\rm I} &=& \hat{y}-\nu_{{\rm I}y}^\prime\cot\tilde{\theta}_{\rm I}\hat{x}^\prime_{\rm I}-\nu_{{\rm I}y}^\prime\hat{z}^\prime_{\rm I},
\\
\hat{z}^{\prime\prime}_{\rm I} &=& \hat{z}^\prime_{\rm I}+\nu_{{\rm I}y}^\prime\hat{y} .
\eea
Now Eq.~\eqref{eq:incident_boost} reads
\bea
\begin{split}
\tilde{\bm{{\cal E}}}_{\rm I}&=\frac{{\cal E}_0}{\sqrt{1+|m|^2}}\left(\hat{x}^{\prime\prime}_{\rm I}
-\nu_{{\rm I}y}^\prime\left(\cot\tilde{\theta}_{\rm I}+\beta\cos\theta_{\rm I}\right)\hat{y}^{\prime\prime}_{\rm I}\right)
\\
&+\frac{m{\cal E}_0}{\sqrt{1+|m|^2}}\left(\nu_{{\rm I}y}^\prime\left(\cot\tilde{\theta}_{\rm I}+\beta\cos\theta_{\rm I}\right)\hat{x}^{\prime\prime}_{\rm I}
+\hat{y}^{\prime\prime}_{\rm I}\right) .
\end{split}
\label{eq:incident_boost2}
\eea
Then the amplitude of the transmitted electric field reads
\be
\begin{split}
\tilde{\bm{{\cal E}}}_{\rm T}&=
\frac{t_p{\cal E}_0}{\sqrt{1+|m|^2}}\left(\hat{x}^{\prime\prime}_{\rm T}
-\frac{t_s}{t_p}\nu_{{\rm I}y}^\prime\left(\cot\tilde{\theta}_{\rm I}+\beta\cos\theta_{\rm I}\right)\hat{y}^{\prime\prime}_{\rm T}\right)
\\
&+\frac{mt_s{\cal E}_0}{\sqrt{1+|m|^2}}\left(\frac{t_p}{t_s}\nu_{{\rm I}y}^\prime\left(\cot\tilde{\theta}_{\rm I}+\beta\cos\theta_{\rm I}\right)\hat{x}^{\prime\prime}_{\rm T}
+\hat{y}^{\prime\prime}_{\rm T}\right) ,
\end{split}
\label{eq:refraction_boost}
\ee
where $t_s$ and $t_p$ are the Fresnel coefficients. 
In terms of the central wave vector coordinates, it reads
\be
\begin{split}
&\tilde{\bm{{\cal E}}}_{\rm T} =
\\
&\frac{t_p{\cal E}_0}{\sqrt{1+|m|^2}}\Bigl(\hat{x}^{\prime}_{\rm T}
\\
&+\left(\frac{\nu_{{\rm I}y}^\prime}{\sin\tilde{\theta}_{\rm I}}\left(\cos\tilde{\theta}_{\rm T}-\frac{t_s}{t_p}\cos\tilde{\theta}_{\rm I}\right)-\frac{t_s}{t_p}\beta\cos\theta_{\rm I}\nu_{{\rm I}y}^\prime\right)\hat{y}\Bigr) ,
\\
+&\frac{mt_s{\cal E}_0}{\sqrt{1+|m|^2}}\Bigl(
\hat{y}-\frac{\nu_{{\rm I}y}^\prime}{n}\hat{z}^\prime_{\rm T}
\\
&-\left(\frac{\nu_{{\rm I}y}^\prime}{\sin\tilde{\theta}_{\rm I}}\left(\cos\tilde{\theta}_{\rm T}-\frac{t_p}{t_s}\cos\tilde{\theta}_{\rm I}\right)-\frac{t_p}{t_s}\beta\cos\theta_{\rm I}\nu_{{\rm I}y}^\prime\right)\hat{x}^{\prime}_{\rm T}\Bigr) ,\;\;\;\;\;\;
\label{eq:refraction_boost2}
\end{split}
\ee
where we used Snell's law: $\sin\tilde{\theta}_{\rm I}=n\sin\tilde{\theta}_{\rm T}$, and $\nu_{{\rm I}y}^\prime=n\nu_{{\rm T}y}^\prime$.
By considering the inverse Lorentz transformation with the boost parameter $\bm\beta=\beta\hat{x}=\beta\cos\tilde{\theta}_{\rm T}\hat{x}^\prime_{\rm T}+\beta\sin\tilde{\theta}_{\rm T}\hat{z}^\prime_{\rm T}$,
we obtain the transmitted wave in the laboratory frame (in a magnetoelectric material) as
\be
\begin{split}
\tilde{\bm{E}}_{\rm T} &=
\\
&\frac{t_pE_0}{\sqrt{1+|m|^2}}\Bigl(\hat{x}_{\rm T}
+\nu_{{\rm I}y}\frac{\cos\tilde{\theta}_{\rm T}-(t_s/t_p)\cos\tilde{\theta}_{\rm I}}{\sin\tilde{\theta}_{\rm I}}\hat{y}\Bigr)
\\
+&
\frac{mt_sE_0}{\sqrt{1+|m|^2}}\Bigl(
\hat{y}-\frac{\nu_{{\rm I}y}}{n}\hat{z}_{\rm T}
-\nu_{{\rm I}y}\frac{\cos\tilde{\theta}_{\rm T}-(t_p/t_s)\cos\tilde{\theta}_{\rm I}}{\sin\tilde{\theta}_{\rm I}}\hat{x}_{\rm T}\Bigr) ,\;\;\;\;\;\;
\end{split}
\label{eq:refraction}
\ee
where 
$\hat{x}_{\rm T}=\hat{\bar{x}}^\prime_{\rm T}+n\beta\cos\tilde{\theta}_{\rm T}\hat{\bar{z}}^\prime_{\rm T}$, and $\hat{z}_{\rm T}=\hat{\bar{z}}^\prime_{\rm T}-n\beta\cos\tilde{\theta}_{\rm T}\hat{\bar{x}}^\prime_{\rm T}$. 
The rotation is needed to satisfy the transversality condition.

Now we compute the transverse displacements of a transmitted wave packet in a magnetoelectric material.
First, we consider a horizontally-polarized incident beam ($m=0$) with finite distribution along the $y$ direction~\cite{Hosten787}:
\bea
\begin{split}
\Psi_{\rm initial}&=\frac{\Psi(y)}{\sqrt{2}}
\begin{pmatrix}
1 \\
1
\end{pmatrix}
\\
&=
\int dp_y e^{ip_yy}\frac{\Phi(p_y)}{\sqrt{2}}
\begin{pmatrix}
1 \\
1
\end{pmatrix} ,
\end{split}
\label{eq:initial}
\eea
where $(\eta_+,\eta_-)^T$ represents the state of polarization in the spin basis.
The right- and left-handed states rotate in an opposite way upon transmission to satisfy the transversality condition when a finite distribution of a wave packet is taken into account~\cite{Hosten787}.  
In real-space coordinates, such rotations result in the transverse displacements of central positions~\cite{Hosten787}:
\bea
\begin{split}
\Psi_{\rm final}&=t_p\int dp_y e^{ip_yy}\frac{\Phi(p_y)}{\sqrt{2}}
\begin{pmatrix}
e^{-i p_y\delta^{\rm H}} \\
e^{+i p_y\delta^{\rm H}}
\end{pmatrix}
 \\
&=\frac{t_p}{\sqrt{2}}
\begin{pmatrix}
\Psi(y-\delta^{\rm H}) \\
\Psi(y+\delta^{\rm H}) 
\end{pmatrix}. 
\end{split}
\label{eq:final}
\eea
Similar arguments hold for a vertical polarization ($m=\infty$) and lead to the transverse displacement shown below.
Form Eq.~\eqref{eq:refraction}, the rotations and displacements of the spin components ($\pm$) for horizontally and vertically polarized lights are given as~\cite{Hosten787}
\bea
\delta^{\rm H}_\pm&=&\pm\frac{\lambda}{2\pi}\frac{\cos\tilde{\theta}_{\rm T}-(t_s/t_p)\cos\tilde{\theta}_{\rm I}}{\sin\tilde{\theta}_{\rm I}} ,
\label{eq:displacement1}
\\
\delta^{\rm V}_\pm&=&\pm\frac{\lambda}{2\pi}\frac{\cos\tilde{\theta}_{\rm T}-(t_p/t_s)\cos\tilde{\theta}_{\rm I}}{\sin\tilde{\theta}_{\rm I}},
\label{eq:displacement2}
\eea
where $\tilde{\theta}_{\rm I}=\theta_{\rm I}+g\cos\theta_{\rm I}/(n^2-1)$, and $\lambda$ is the wave length of a light in incident medium.
The Fresnel coefficients $t_s$ and $t_p$, and the refraction angle $\tilde{\theta}_{\rm T}$ are given explicitly as
\bea
&&t_s 
=\frac{2\mu \cos\tilde{\theta}_{\rm I}}{\mu \cos\tilde{\theta}_{\rm I}+n \cos\tilde{\theta}_{\rm T}}  ,
\label{eq:ts}
\\
&&t_p 
=\frac{2\mu \cos\tilde{\theta}_{\rm I}}{n \cos\tilde{\theta}_{\rm I}+\mu \cos\tilde{\theta}_{\rm T}}  ,
\label{eq:tp}
\\
&&\sin\tilde{\theta}_{\rm T} =\frac{\sin\tilde{\theta}_{\rm I}}{n} .
\label{eq:refracted_angle}
\eea
The transverse displacements of a refracted wave packet due to the chiral magnetic effect at a surface of magnetoelectric materials can be computed as those in the ``rest" frame due to the spin Hall effect with the modified incident angle $\tilde{\theta}_{\rm I}$.
The incident angle dependence of $\delta^{\rm H}_+$ and $\delta^{\rm V}_+$ is discussed in the main text.

\bibliography{./photon}